\documentclass[aps,prl,twocolumn]{revtex4}% PRL
\usepackage{amsmath,amssymb,amsthm,mathrsfs,amsfonts,dsfont} 
\usepackage{graphicx}
\usepackage{subfigure}
\usepackage{amsbsy}
\usepackage{bm}

\def\id{{\mathds{1}}}
\def\bs{\bm{s}}
\def\bx{\boldsymbol{\xi}}
\def\X{\mathbf{X}}

\begin{document}

\title{Associative memory model with arbitrary Hebbian length}
\author{Zijian Jiang}
\thanks{Equal contribution.}
\affiliation{PMI Lab, School of Physics,
Sun Yat-sen University, Guangzhou 510275, People's Republic of China}
\author{Jianwen Zhou}
\thanks{Equal contribution.}
\affiliation{PMI Lab, School of Physics,
Sun Yat-sen University, Guangzhou 510275, People's Republic of China}
\author{Tianqi Hou}
\thanks{Equal contribution.}
\affiliation{Department of Physics, the Hong Kong University of Science and Technology, Clear Water Bay, Hong Kong}
\affiliation{Theory Lab, Central Research Institute, 2012 Labs, Huawei Technologies Co., Ltd.}
\author{K. Y. Michael Wong}
\affiliation{Department of Physics, the Hong Kong University of Science and Technology, Clear Water Bay, Hong Kong}
\author{Haiping Huang}
\email{huanghp7@mail.sysu.edu.cn}
\affiliation{PMI Lab, School of Physics,
Sun Yat-sen University, Guangzhou 510275, People's Republic of China}
\date{\today}

\begin{abstract}
Conversion of temporal to spatial correlations in the cortex is one of the most intriguing functions in
the brain. The learning at synapses
triggering the correlation conversion
can take place in a wide integration window, whose influence on the correlation conversion remains elusive. Here, we propose a generalized 
associative memory model with arbitrary Hebbian length. The model can be analytically solved, and predicts that a small Hebbian length can already significantly
enhance the correlation conversion, i.e., the stimulus-induced attractor can be highly correlated with a significant number of patterns in the stored sequence,
thereby facilitating state transitions in the neural representation space. Moreover, an anti-Hebbian component is able to reshape the energy landscape of memories, akin to the function of sleep.
Our work thus establishes the fundamental connection between associative memory, Hebbian length, and correlation conversion in the brain.
\end{abstract}

%\pacs{02.50.Tt, 87.19.L-, 75.10.Nr}
 \maketitle

%%%%%%%%%%%%%%%%%%%%%%%%%%%%%%%%%%%%%%%%%%%%%%%%%%%%%%%%%%%%%%%%%
\textit{Introduction.}
Associative learning and memory is one fundamental brain function across many species including rodents and primates~\cite{Guzman-2016,Fusi-2020}.
The standard Hopfield network, based on Hebbian learning rules, establishes a seminal model to explore rich properties of 
associative memory in both artificial and biological neural networks~\cite{Hopfield-1982,Amari-1977}. As a classic example,
the monkey's temporal cortex was observed to be able to convert the temporal correlation of stimuli into
the spatial correlation in neural activity~\cite{Miya-1988a,Miya-1988b}, which can be modeled by considering Hebbian interactions among 
neighboring random independent patterns~\cite{Amit-1993}. 
For an external stimulus being part of temporally ordered sequence, the elicited neural activity has
a correlation with neighboring patterns of the sequence which decays until vanishing at a finite separation of the patterns.
This correlated attractor phase is in contrast to the Hopfield model where all attractors corresponding to the stored patterns are all
uncorrelated fixed points in the network dynamics. A recent study argued that the combination of Hebbian and anti-Hebbian learning can significantly 
increase the span of the temporal association~\cite{Fukai-2019}. However, wide learning windows of various widths have been observed in 
biological synaptic plasticity~\cite{Bittner-2017,Hebbian-2018,Hebbian-2020}. Whether this microscopic temporal correlation in synaptic learning 
affects the global behavior of correlated attractors remains therefore elusive. Hence, 
a full understanding of how the temporal correlation among stimuli evokes the spatially correlated
neural activity
is still lacking.

Here, we propose a theoretical model of associative memory with arbitrary Hebbian length, corresponding to wide learning windows.
This model can be analytically solved, providing us exact mechanisms underlying the correlated attractor phase.
In particular, we find that even with only Hebbian learning, the wide learning window can give rise to 
a large correlation span, which suggests a distinct synaptic mechanism from that argued in the recent work~\cite{Fukai-2019}. Most importantly,
our model reveals that an anti-Hebbian learning for the non-concurrent patterns could reshape the energy landscape, removing irrelevant 
attractors, which may be related to the hypothesis of unlearning effects in rapid-eye-movement sleep (e.g., get rid of unimportant memory)~\cite{Crick-1983,Hopfield-1983,Sleep-2010,Zhou-2020rem}.

\textit{Model.}
In this study, we construct an associative memory model by the Hebbian learning~\cite{Hebb-1949}, which shapes the coupling strength between two neurons.
We assume that all $N$ neurons are fully-connected without self-interactions, which constructs an associative memory of
$P$ random patterns ($\bx$). These patterns form a cyclic sequence, corresponding to a
repeated presentation of an ordered sequence of independent items in monkey experiments
~\cite{Miya-1988a,Miya-1988b}. Therefore, the coupling matrix of the associative memory model
can be specified as follows,
\begin{equation}
 \label{coupling}
 J_{ij}=\frac{1}{N}\sum_{\mu=1}^P\left[c\xi_i^\mu\xi_j^\mu+\gamma\sum_{r=1}^d(\xi_i^{\mu+r}\xi_j^\mu+\xi_i^\mu\xi_j^{\mu+r})\right],
\end{equation}
where $c$ specifies the standard Hebbian strength (concurrent Hebbian terms), $\gamma$ specifies the coupling strength between $r$-separated patterns (non-concurrent Hebbian terms), and $d$ is thus 
the Hebbian length of our model. The case of $d=1$ has been studied by previous works~\cite{Amit-1993,Tsodyks-1994,Fukai-2019}, while $d=0$ recovers the standard 
Hopfield model~\cite{Amari-1977,Hopfield-1982,Amit-1985}. Setting an arbitrary $d$ corresponds to potential wide learning windows observed 
in neural circuits~\cite{Hebbian-1999,Abbott-2000,Tri-2006,Bittner-2017,Hebbian-2018,Hebbian-2020}. For simplicity, $P(\xi^\mu_i=\pm1)=1/2$ for each pair $(i,\mu)$. 

The coupling is symmetric, and thus an equilibrium state $\bs$ exists, captured by
the following Boltzmann distribution,
\begin{equation}
 P(\bs)=\frac{1}{Z}e^{-\beta\mathcal{H}(\bs)},
\end{equation}
where $\mathcal{H}(\bs)=-\frac{1}{2}\sum_{i\neq j}J_{ij}s_is_j$ being the Hamiltonian, $\beta$ is an inverse temperature,
and $Z$ is the pattern-dependent partition function.
Note that we can re-arrange the coupling matrix as $\mathbf{J}=\frac{1}{N}\bx^{\rm T}\mathbf{X}\bx$, where the circulant matrix $X_{\mu\nu}$ is introduced as follows~\cite{Gray-2005},
\begin{equation}
 X_{\mu\nu}=c\delta_{\mu\nu}+\gamma\sum_{r=1}^d(\delta_{\mu,(\nu-r)\,{\rm mod}\,P}+\delta_{\mu,(\nu+r)\,{\rm mod}\,P}).
\end{equation}
Then, the Hamiltonian can be expressed as $\mathcal{H}(\bs)=-\frac{N}{2}\mathbf{M}^{\rm T}\mathbf{X}\mathbf{M}$, where $\mathbf{M}$ denotes the pattern-state overlap vector
whose component $M^\mu=\frac{1}{N}\sum_i\xi_i^\mu s_i$.

Like in the standard Hopfield model~\cite{Amari-1977,Hopfield-1982}, the state of each neuron is determined by its local field $h_i$, which can be written as
$h_i=\sum_jJ_{ij}s_j$.
By inserting the coupling matrix, we get a new expression,
\begin{equation}
 h_i=\sum_{\mu}\xi_i^\mu\left(cM^\mu+\gamma\sum_{r=1}^d(M^{\mu-r}+M^{\mu+r})\right).
\end{equation}
Due to the statistical independence of the patterns, the overlap has a mean-field expression~\cite{SM},
\begin{equation}\label{mf-olap}
 M^\mu=\left\langle\xi^\mu\operatorname{sgn}\left(\sum_{\nu}\xi^\nu\left(cM^\nu+\gamma\sum_{r=1}^d(M^{\nu-r}+M^{\nu+r})\right)\right)\right\rangle,
\end{equation}
where $\langle\cdot\rangle$ denotes the disorder average over the pattern, and the zero-temperature limit ($\beta\to\infty$) is considered.
In this limit, the dynamics is noiseless, and for $d=1$ the overlap with the pattern used as a stimulus displays a largest value and was found to decay symmetrically until vanishing at
a pattern-separation distance of five~\cite{Amit-1993}, which is independent of the number of patterns $P$. This shows that, although the patterns are uncorrelated,
the retrieved attractor starting from the stimulus has macroscopically significant
overlaps with neighboring patterns within a finite distance.
We call this kind of attractor correlated attractor. 

In the same spirit, the correlation of activities in two attractors can be computed as
\begin{equation}\label{mf-corr}
 C(\mu,\mu')=\langle\operatorname{sgn}(h^\mu)\operatorname{sgn}(h^{\mu'})\rangle,
\end{equation}
where $h^\mu=\sum_{\nu}\xi^\nu\left(cM^\nu_\mu+\gamma\sum_{r=1}^d(M^{\nu-r}_\mu+M^{\nu+r}_\mu)\right)$~\cite{SM}, and $M^\nu_\mu$ defines the overlap of the attractor
corresponding to the stimulus $\mu$ with the pattern number $\nu$.
The behavior of $C(\mu,\mu')$ shows the emergence of correlated attractors from a network storing uncorrelated patterns.
This attractor correlation decays with the separation of the patterns in the sequence from the stimulus pattern, 
where we can determine the critical distance (correlation length denoted as $\ell_c$) beyond which the correlation value 
falls below $10^{-2}$. This captures the basic coding strategies in the temporal cortex of the monkey, which 
is able to convert the temporal correlation among visual stimuli into a spatial correlation in the sustained neural activities evoked by 
the stimuli~\cite{Miya-1988a,Miya-1988b,Amit-1993}.
It is thus interesting to explore analytically how the Hebbian length (or other model 
parameters) affects properties of the correlated attractor.

\textit{A statistical mechanics analysis.}
Now we calculate the free energy of the model for the extensive-load case $\alpha=P/N\sim\mathcal{O}(1)$. To derive a typical behavior of the model, we need to perform a disorder average of 
$\ln Z$, which can be tackled by the replica method: $-\beta f=\lim_{n\to 0,N\to\infty}\frac{\ln\langle Z^n\rangle}{nN}$ (e.g., see~\cite{Huang-2019,Huang-2020}). In essence, $n$ copies of the original
system are introduced. The calculation details are given
in the Supplemental Material~\cite{SM}. In accord with the aforementioned noiseless dynamics, we 
are interested in the zero-temperature phase diagram. The finite-temperature analysis is straightforward~\cite{SM}.

\begin{figure*}
\centering
     \includegraphics[bb=6 5 855 574,scale=0.5]{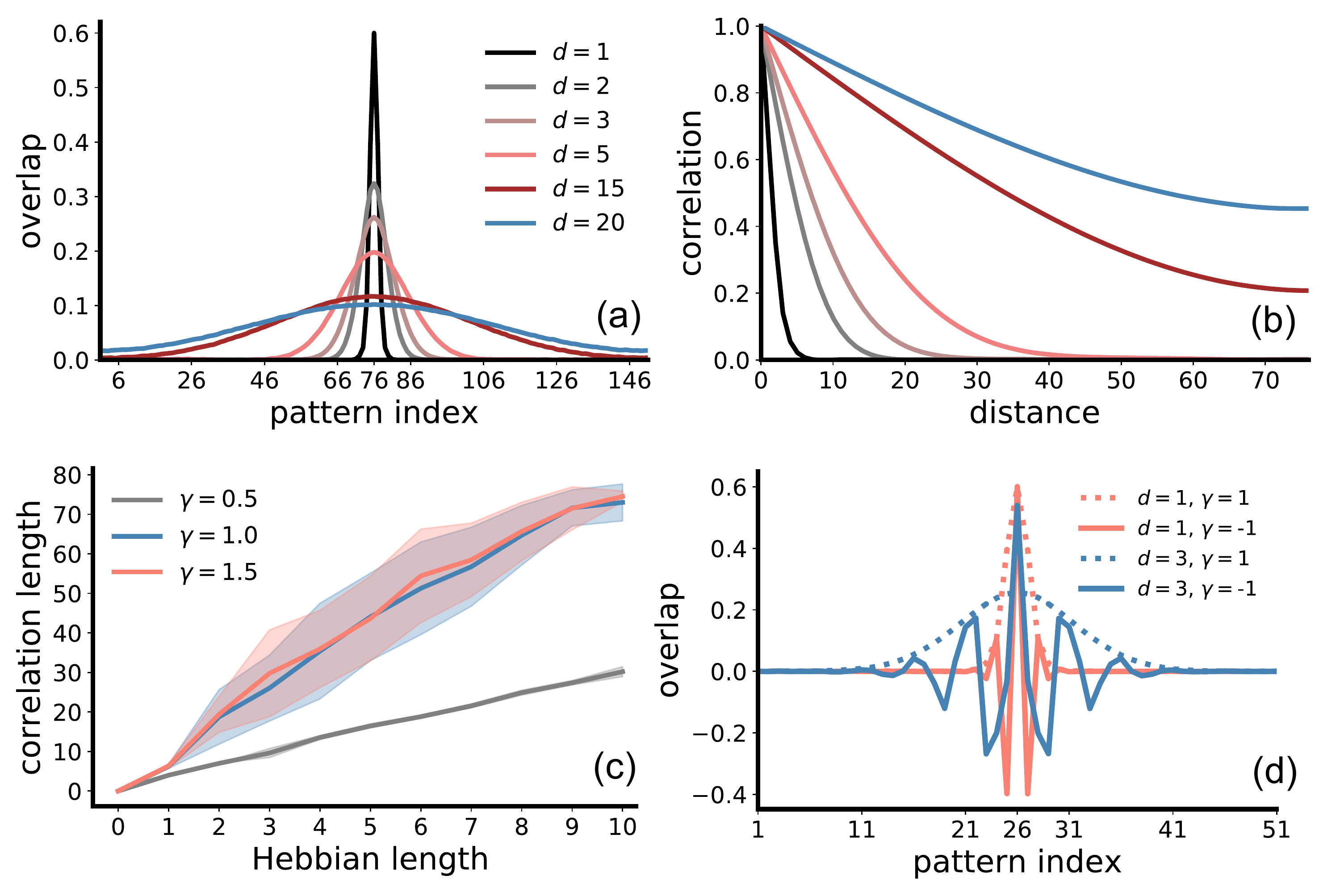}
  \caption{
  (Color online) Transforming temporal to spatial correlations with arbitrary Hebbian length.
    (a) Overlap profile with varying Hebbian length $d$. Other model parameters are $P=151$, and $c=1.0$ and $\gamma = 1.0$. 
    (b) Correlation between attractors versus their distance. The distance is defined as the separation from the corresponding stimulating patterns in the cyclic sequence.
    Other settings are the same as in (a). 
    (c) Correlation length versus Hebbian length $d$. Other parameters are $P=151$, and $c=1.0$. 
    The correlation length $\ell_c=\min\{\ell|C(\ell=|\mu-\mu'|)<10^{-2}\}-1$. Fluctuations are the standard errors
    calculated from $30$ trials. (d) Negative $\gamma$ leads to the oscillatory behavior of the overlap profile. Other parameters 
    are $P=51$, and $c=1.0$.
  }\label{overlap}
\end{figure*}

The analysis of the $n$ replicas leads to the order parameters $M_a^\mu=\frac{1}{N}\sum_i^\mu\xi_i^\mu s_i^a$ and the state overlap $q_{ab}=\frac{1}{N}\sum_is_i^as_i^b$.
For simplicity, we take the replica symmetric assumption~\cite{Tsodyks-1994}, where the order parameters ($\{M^\mu_a,q_{ab}\}$)
and their conjugate counterparts ($\{\hat{M}_a^\mu,\hat{q}_{ab}\}$) do not depend on the replica index ($a$ or $b$).
 The thermodynamic limit makes a saddle point analysis of the free energy reasonable, which leads to the following saddle-point equations:
\begin{subequations}\label{sdeam}
 \begin{align}
  M^\mu&=\left\langle\xi^\mu\operatorname{erf}\left(\frac{\sum_{\nu=1}^S\hat{M}^\nu\xi^\nu}{\sqrt{2\hat{q}}}\right)\right\rangle,\\
  \hat{M}^\mu&=[\mathbf{K}\mathbf{M}]_\mu,\\
  R&=\sqrt{\frac{2}{\pi\hat{q}}}\left\langle \exp\left(-\frac{\left(\sum_{\nu=1}^S\hat{M}^\nu\xi^\nu\right)^2}{2\hat{q}}\right)\right\rangle,\\
  \hat{q}&=\alpha\int_0^1du\frac{\Lambda(u)^2}{(1-R\Lambda(u))^2}+\mathbf{M}^{\rm T}\frac{\partial\mathbf{K}}{\partial R}\mathbf{M},
 \end{align}
\end{subequations}
where $S$ denotes the number of condensed patterns (i.e., $M^\mu$ does not vanish as $N\to\infty$), and $\Lambda(u)=c+2\gamma\sum_{r=1}^d\cos(2\pi ur)$ is the eigenvalue of the matrix $\X$
in the large $P$ limit.
$S$ can be larger than one due to the emergence of the correlated-attractor phase.
$\mathbf{K}$ is an $S\times S$ matrix given by $\mathbf{F}+R^{-1}\id$, where $\id$ is an identity matrix, and $(\mathbf{F}^{-1})_{ij}=w_{j-i}$ being a Toeplitz matrix~\cite{Gray-2005},
whose components ($w_k$) depend on both $R$ and $\Lambda$~\cite{SM}. 
In the zero-temperature limit, $q\to1$, and thus we denote $R=\beta(1-q)$.

For the standard Hopfield model, $\X=\id$, $\Lambda(u)=1$, and thus Eq.~(\ref{sdeam}) reduces to the mean-field equation derived in the seminal work~\cite{Amit-1985}.
In our current setting, the Hebbian length affects both $\hat{q}$ and $\mathbf{K}$ in a highly
nontrivial way. We thus expect the corresponding influence on the global behavior of correlation conversion. 
%%%%%%%%%%%%%%%%%%%%%%%%%%%%%%%%%%%%%%%%%%

%%%%%%%%%%%%%%%%%%%%%%%%%%%%%%%%%%%%%%%%%%%%%%%%%%%%

\textit{Results.}
%%%%%%%%%%%%%%%%%%%%%%%%%%%%%%%%
 We first study the mean field dynamics [Eqs.~(\ref{mf-olap}-\ref{mf-corr})] of the overlap function at finite values of $P$, focusing on impacts of different model parameters.
 As shown in Fig.~\ref{overlap}, increasing the Hebbian length lowers down the peak value of the overlap with the stimulus pattern, and meanwhile, the overlap
 with neighboring patterns grows, thereby making the overlap profile broader. Surprisingly, by increasing the Hebbian length up to only $d=2$, the correlation 
 span is increased by quite a large margin (from $\ell_c=5$ when $d=1$ to $\ell_c=15$ when $d=2$). Compared to fine tuning the (negative) strength of the concurrent
 Hebbian terms~\cite{Fukai-2019}, increasing the Hebbian length is simple and moreover biologically intuitive, as the Hebbian length corresponds to the size of the learning integration window,
 widely observed in neural circuits~\cite{Hebbian-1999,Abbott-2000,Tri-2006,Bittner-2017,Hebbian-2018,Hebbian-2020}. In particular, a large value of $d$ allows for associations of patterns (stimuli) distant with each other
 in the sequence [Fig.~\ref{overlap} (a,b)].
 Interestingly, the overlap profile of $c=-1$ and $d=1$ is exactly the same with that of $c=1$ and $d=2$~\cite{SM}. Furthermore, it requires only $d=15$ for
 the correlation to expand to all patterns in the sequence, for $P=151$ in Fig.~\ref{overlap} (a). In other words, a small value of $d$ can significantly amplify the correlation span [Fig.~\ref{overlap}(c)].
 The corresponding influence of $d$ is tuned by the Hebbian strength $\gamma$, and a large value of $\gamma$ has a less impact on the tuning.
 
 Therefore, our model with arbitrary Hebbian length provides a simple alternative way to control the correlation span of the stimulus-induced attractor,
 which is related to the conversion of the temporal correlations in the stored sequence into the spatial correlations of the neural activities.
 The correlated attractor phase is able to accelerate the transition between two highly correlated attractors (e.g., memories), since both attractors share a large number of 
 common active neurons in their neural representations.
 
 \begin{figure*}
     \includegraphics[bb=4 8 755 327,scale=0.65]{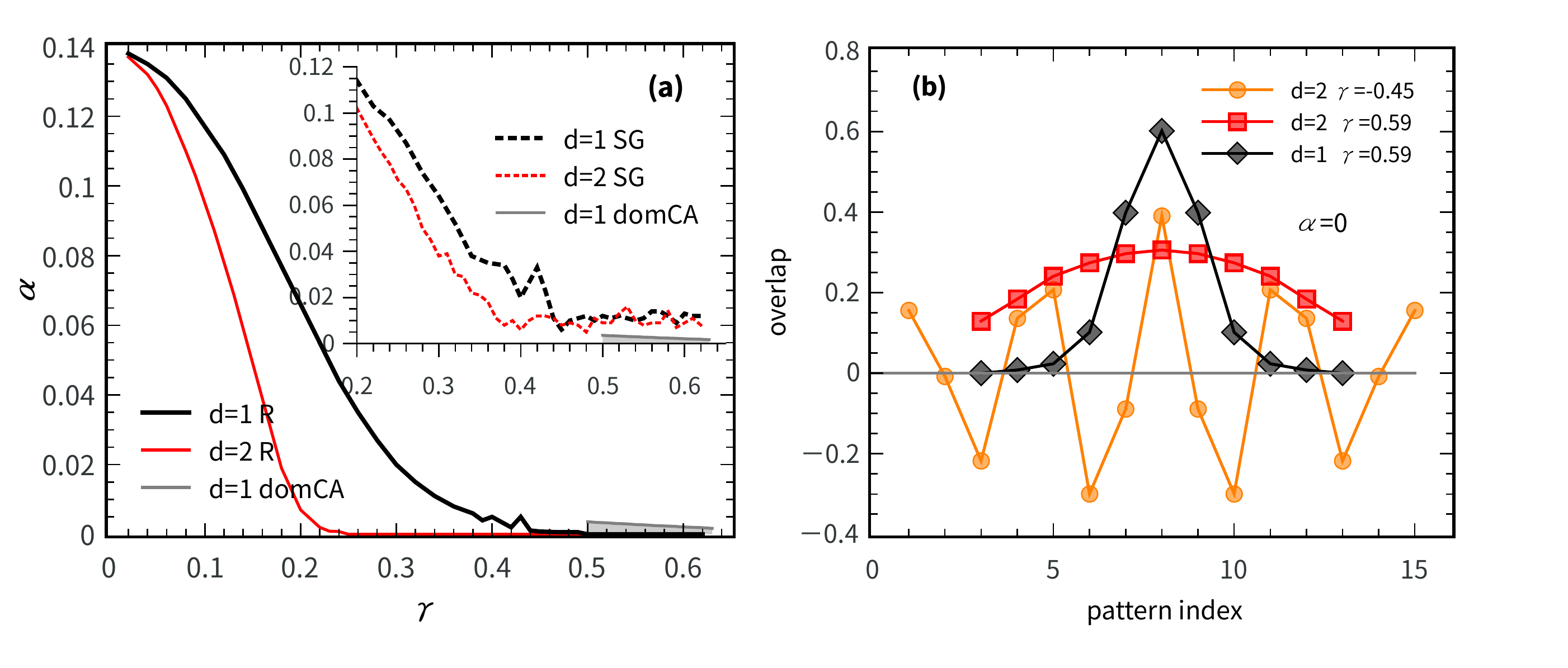}
  \caption{
  (Color online) Phase diagram of the associative memory model in the $(\alpha,\gamma)$ plane given $c=1$.
  (a) The phase boundary shown by the lines delimits the retrieval (R) phase from the region where
  the correlated-attractor (CA) and spin glass (SG) phases compete with each other (above the boundary). 
  The boundary is the condition on which the retrieval phase loses its metastability from below. All shown transitions are of the discontinuous type.
  When $\alpha=0$, the transition point is given by $\gamma_c=0.5$ for $d=1$, while $\gamma_c=0.25$ for $d=2$. The inset
  shows the boundary line above which the spin glass phase is dominant. Note that for $d=1$, there exists a very narrow regime (indicated by the shadow)
  within which the correlated-attractor phase is dominant (domCA). (b) Overlap profiles obtained from the statistical mechanics theory.
  All overlap profiles are defined as in Fig.~\ref{overlap}, and obtained by solving the saddle-point equation of the model when $\alpha=0$ and $d=2$ (or $d=1$).
  All theoretical results are obtained by assuming that $S=11$, except that for negative values of $\gamma$, we use $S=15$. Note that the results are not sensitive to
  the value of $S$ (e.g., $S=11$ or $S=13$).
  }\label{transition}
\end{figure*}

 Next, we explore the effect of the non-concurrent anti-Hebbian terms. These terms are characterized by negative values of $\gamma$, which competes with the concurrent Hebbian terms ($c>0$).
 In addition, the anti-Hebbian terms correspond to the unlearning process introduced to verify the hypothesis of memory consolidation or erasure in sleep~\cite{Hopfield-1983,Crick-1983,Sleep-2010}.
 Here, we find that the non-concurrent anti-Hebbian terms remove some specified attractors, which appears in the original energy landscape of the model without anti-Hebbian effects.
 In contrast, the corresponding sign-reversed attractors are preferred, indicated by the negative overlaps in Fig.~\ref{overlap} (d). This
 interesting observation could be explained by the energy landscape in terms of overlap functions. We recast the Hamiltonian as $\mathcal{H}(\mathbf{M})=-\frac{Nc}{2}\sum_{\mu}(M^\mu)^2-N\gamma\sum_{\mu}\sum_{r=1}^dM^\mu M^{\mu+r}$,
 where the first Hebbian term is always negative ($c>0$), while the second term ($\gamma<0$) requires that some specific overlap with a particular pattern index must take a 
 negative value for a lower energy. In other words, the unlearning terms are able to reshape the energy landscape, by consolidating some memories while erasing other memories, akin to
 the function of both types of sleep: the rapid eye movement (REM) sleep is hypothesized to remove unnecessary memories while the slow wave sleep contributes to the consolidation of important 
 memories~\cite{Sleep-2010,Sleep-2017}.

 Finally, we look at the phase diagram. We consider only $d=1$ and $d=2$. Other values of $d$ could be analogously studied with our theory.
 As shown in Fig.~\ref{transition} (a), we identify three phases.
 One is the retrieval phase where only one overlap component is of the order one, i.e.,
 $M^\mu=m\delta_{\mu\nu}$, where $\nu$ indicates the stimulating pattern. Given the value of $\alpha$, increasing the value of $\gamma$ would finally make the retrieval phase lose its metastability, after which 
 the correlated-attractor phase becomes metastable. The line separating these two phases is thus 
 the first-order transition. The correlated-attractor phase is characterized by the stimulus-induced attractors being highly correlated with a finite number of patterns in the stored sequence.
 In other words, the value of the corresponding overlap decays with the distance between the patterns in the sequence and the one used as the stimulus.
 The numerical solutions of the saddle-point equations obtained by the replica theory [Eq.~(\ref{sdeam})] reproduce the key features of the mean-field dynamics of the overlap [Fig.~\ref{overlap}, and Fig.~\ref{transition} (b)],
 which corresponds to $\alpha=0$ in our theory. 
 
 Our theory predicts that the value of $d$ can be used to expand the correlation span of
 the correlated-attractor, and moreover reshape significantly the phase diagram. When $\alpha=0$, the threshold for the
 dominant retrieval phase is $\gamma_c=0.5$ for $d=1$, but $\gamma_c=0.25$ for $d=2$. In the presence of a finite $\alpha$,
 the retrieval phase loses its metastability at a smaller value of $\gamma$ for $d=2$ than for $d=1$ [Fig.~\ref{transition} (a)]. After that,
 the spin glass phase characterized by $M^\mu=0$ ($\forall\mu$) appears and competes with the correlated attractor phase, until the point where
 the spin glass phase becomes dominant (global minimum of the free energy), as shown in the inset of Fig.~\ref{transition} (a).
 Remarkably, for $d=1$, we identify a narrow regime for $\gamma>0.5$ [the shadow in Fig.~\ref{transition} (a)], where the correlated-attractor phase becomes dominant. This regime shrinks gradually as
 $\gamma$ increases. If noisy neural dynamics is allowed (e.g., at a non-zero temperature), the spin glass phase would be replaced by
 a paramagnetic phase at a continuous transition (see a detailed exploration in an accompany paper~\cite{Huang-2021b}).
 This transition line is also strongly affected by the Hebbian length.

In particular, our theoretical analysis also reproduces the unlearning effects observed in the 
mean field dynamics. Furthermore, a critical strength of $\gamma_c=-0.25$ for the oscillatory phase is predicted for $d=2$.
$\gamma_c=-0.5$ for $d=1$.
When $\gamma<\gamma_c$, the unlearning effect of non-concurrent anti-Hebbian terms takes place, preferring some particular patterns rather than their sign-reversed counterparts.
In other words, the (spin reversal) symmetry in the Hamiltonian is broken, and the negative $\gamma$ selects particular patterns, which suggests that the energy landscape is reshaped, and further the information storage is re-optimized~\cite{Barra-2019,Dot-1991,Nokura-1998}.
This intriguing phenomenon thus establishes the connection between the Hebbian length, anti-Hebbian effect, and memory function of unlearning.

\textit{Conclusion.}
%%%%%%%%%%%%%%%%%%%%%%%%%%%%%%%%%%%%%%%%%%%%%%%%%%%%%%%%%
In this Letter, we propose the associative memory model of arbitrary Hebbian length, which considers both the wide learning integration window and temporal-to-spatial correlation conversion observed in the brain.
Our theory predicts that a small value of Hebbian length (e.g., $d=2$) can significantly expand the correlation span of the stimulus-induced attractors. Therefore, it seems unnecessary to fine tune the concurrent Hebbian strength $c$.
Instead, by increasing $d$ only one can achieve the same goal of enhanced spatial correlations in neural attractors.
Moreover, a negative value of $\gamma$ can trigger an oscillatory behavior of the overlap profile, removing some irrelevant pattern attractors in the energy landscape, thereby playing the role of 
regulating the stored memories. Lastly, the Hebbian length could change strongly the phase diagram of the model. Increasing slightly the value of $d$ would significantly suppress the retrieval phase, and
moreover strongly affect the metastable regime of the correlation conversion. Taken together, our theory of the generalized associative memory model provides insights about the interplay
between three important concepts---arbitrary Hebbian length, unlearning, and correlation conversion in 
neural circuits.

The encoding of pattern sequences in correlated attractors is reminiscent of encoding a continuous sequence of patterns in 
continuous attractor neural networks, which are useful for processing continuous information~\cite{Fung-2010,Monasson-2020}.

%%%%%%%%%%%%%%%%%%%%%%%%%%%%%%%%%%%%%%%%%%%%%%%%%%%%%%%
%\section*{Acknowledgments}
\begin{acknowledgments}
This research was supported by the National Natural Science Foundation of China for
Grant No. 11805284 (HH) and the start-up budget 74130-18831109 of the 100-talent-program 
of Sun Yat-sen University (HH), and research grants council of Hong Kong (grant numbers 16302419 and 16302619) (MW). 
\end{acknowledgments}
%%%%%%%%%%%%%%%%%%%%%%%%%%%%%%%%%%%%%%%%%%%%%%%%%%%%%%%%%%%%%%%

%\bibliography{ref}

\begin{thebibliography}{33}
\expandafter\ifx\csname natexlab\endcsname\relax\def\natexlab#1{#1}\fi
\expandafter\ifx\csname bibnamefont\endcsname\relax
  \def\bibnamefont#1{#1}\fi
\expandafter\ifx\csname bibfnamefont\endcsname\relax
  \def\bibfnamefont#1{#1}\fi
\expandafter\ifx\csname citenamefont\endcsname\relax
  \def\citenamefont#1{#1}\fi
\expandafter\ifx\csname url\endcsname\relax
  \def\url#1{\texttt{#1}}\fi
\expandafter\ifx\csname urlprefix\endcsname\relax\def\urlprefix{URL }\fi
\providecommand{\bibinfo}[2]{#2}
\providecommand{\eprint}[2][]{\url{#2}}

\bibitem[{\citenamefont{{Guzman} et~al.}(2016)\citenamefont{{Guzman},
  {Schlogl}, {Frotscher}, and {Jonas}}}]{Guzman-2016}
\bibinfo{author}{\bibfnamefont{S.~J.} \bibnamefont{{Guzman}}},
  \bibinfo{author}{\bibfnamefont{A.}~\bibnamefont{{Schlogl}}},
  \bibinfo{author}{\bibfnamefont{M.}~\bibnamefont{{Frotscher}}},
  \bibnamefont{and} \bibinfo{author}{\bibfnamefont{P.}~\bibnamefont{{Jonas}}},
  \bibinfo{journal}{Science} \textbf{\bibinfo{volume}{353}},
  \bibinfo{pages}{1117} (\bibinfo{year}{2016}).

\bibitem[{\citenamefont{{Ahmed} et~al.}(2020)\citenamefont{{Ahmed},
  {Priestley}, {Castro}, {Stefanini}, {Canales}, {Balough}, {Lavoie},
  {Mazzucato}, {Fusi}, and {Losonczy}}}]{Fusi-2020}
\bibinfo{author}{\bibfnamefont{M.~S.} \bibnamefont{{Ahmed}}},
  \bibinfo{author}{\bibfnamefont{J.~B.} \bibnamefont{{Priestley}}},
  \bibinfo{author}{\bibfnamefont{A.}~\bibnamefont{{Castro}}},
  \bibinfo{author}{\bibfnamefont{F.}~\bibnamefont{{Stefanini}}},
  \bibinfo{author}{\bibfnamefont{A.~S.~S.} \bibnamefont{{Canales}}},
  \bibinfo{author}{\bibfnamefont{E.~M.} \bibnamefont{{Balough}}},
  \bibinfo{author}{\bibfnamefont{E.}~\bibnamefont{{Lavoie}}},
  \bibinfo{author}{\bibfnamefont{L.}~\bibnamefont{{Mazzucato}}},
  \bibinfo{author}{\bibfnamefont{S.}~\bibnamefont{{Fusi}}}, \bibnamefont{and}
  \bibinfo{author}{\bibfnamefont{A.}~\bibnamefont{{Losonczy}}},
  \bibinfo{journal}{Neuron} \textbf{\bibinfo{volume}{107}},
  \bibinfo{pages}{283} (\bibinfo{year}{2020}).

\bibitem[{\citenamefont{Hopfield}(1982)}]{Hopfield-1982}
\bibinfo{author}{\bibfnamefont{J.~J.} \bibnamefont{Hopfield}},
  \bibinfo{journal}{Proceedings of the National Academy of Sciences}
  \textbf{\bibinfo{volume}{79}}, \bibinfo{pages}{2554} (\bibinfo{year}{1982}).

\bibitem[{\citenamefont{Amari}(1977)}]{Amari-1977}
\bibinfo{author}{\bibfnamefont{S.-i.} \bibnamefont{Amari}},
  \bibinfo{journal}{Biological cybernetics} \textbf{\bibinfo{volume}{26}},
  \bibinfo{pages}{175} (\bibinfo{year}{1977}).

\bibitem[{\citenamefont{Miyashita}(1988)}]{Miya-1988a}
\bibinfo{author}{\bibfnamefont{Y.}~\bibnamefont{Miyashita}},
  \bibinfo{journal}{Nature} \textbf{\bibinfo{volume}{335}},
  \bibinfo{pages}{817} (\bibinfo{year}{1988}).

\bibitem[{\citenamefont{Miyashita and Chang}(1988)}]{Miya-1988b}
\bibinfo{author}{\bibfnamefont{Y.}~\bibnamefont{Miyashita}} \bibnamefont{and}
  \bibinfo{author}{\bibfnamefont{H.}~\bibnamefont{Chang}},
  \bibinfo{journal}{Nature} \textbf{\bibinfo{volume}{331}}, \bibinfo{pages}{68}
  (\bibinfo{year}{1988}).

\bibitem[{\citenamefont{Griniasty et~al.}(1993)\citenamefont{Griniasty,
  Tsodyks, and Amit}}]{Amit-1993}
\bibinfo{author}{\bibfnamefont{M.}~\bibnamefont{Griniasty}},
  \bibinfo{author}{\bibfnamefont{M.~V.} \bibnamefont{Tsodyks}},
  \bibnamefont{and} \bibinfo{author}{\bibfnamefont{D.~J.} \bibnamefont{Amit}},
  \bibinfo{journal}{Neural Computation} \textbf{\bibinfo{volume}{5}},
  \bibinfo{pages}{1} (\bibinfo{year}{1993}).

\bibitem[{\citenamefont{{Haga} and {Fukai}}(2019)}]{Fukai-2019}
\bibinfo{author}{\bibfnamefont{T.}~\bibnamefont{{Haga}}} \bibnamefont{and}
  \bibinfo{author}{\bibfnamefont{T.}~\bibnamefont{{Fukai}}},
  \bibinfo{journal}{Physical Review Letters} \textbf{\bibinfo{volume}{123}},
  \bibinfo{pages}{78101} (\bibinfo{year}{2019}).

\bibitem[{\citenamefont{Bittner et~al.}(2017)\citenamefont{Bittner, Milstein,
  Grienberger, Romani, and Magee}}]{Bittner-2017}
\bibinfo{author}{\bibfnamefont{K.~C.} \bibnamefont{Bittner}},
  \bibinfo{author}{\bibfnamefont{A.~D.} \bibnamefont{Milstein}},
  \bibinfo{author}{\bibfnamefont{C.}~\bibnamefont{Grienberger}},
  \bibinfo{author}{\bibfnamefont{S.}~\bibnamefont{Romani}}, \bibnamefont{and}
  \bibinfo{author}{\bibfnamefont{J.~C.} \bibnamefont{Magee}},
  \bibinfo{journal}{Science} \textbf{\bibinfo{volume}{357}},
  \bibinfo{pages}{1033} (\bibinfo{year}{2017}).

\bibitem[{\citenamefont{Gerstner et~al.}(2018)\citenamefont{Gerstner, Lehmann,
  Liakoni, Corneil, and Brea}}]{Hebbian-2018}
\bibinfo{author}{\bibfnamefont{W.}~\bibnamefont{Gerstner}},
  \bibinfo{author}{\bibfnamefont{M.}~\bibnamefont{Lehmann}},
  \bibinfo{author}{\bibfnamefont{V.}~\bibnamefont{Liakoni}},
  \bibinfo{author}{\bibfnamefont{D.}~\bibnamefont{Corneil}}, \bibnamefont{and}
  \bibinfo{author}{\bibfnamefont{J.}~\bibnamefont{Brea}},
  \bibinfo{journal}{Frontiers in Neural Circuits}
  \textbf{\bibinfo{volume}{12}}, \bibinfo{pages}{53} (\bibinfo{year}{2018}).

\bibitem[{\citenamefont{Reifenstein and Kempter}(2020)}]{Hebbian-2020}
\bibinfo{author}{\bibfnamefont{E.~T.} \bibnamefont{Reifenstein}}
  \bibnamefont{and} \bibinfo{author}{\bibfnamefont{R.}~\bibnamefont{Kempter}},
  \bibinfo{journal}{bioRxiv}  (\bibinfo{year}{2020}),
  \urlprefix\url{https://www.biorxiv.org/content/early/2020/04/15/2020.04.13.039826}.

\bibitem[{\citenamefont{Crick and Mitchison}(1983)}]{Crick-1983}
\bibinfo{author}{\bibfnamefont{F.}~\bibnamefont{Crick}} \bibnamefont{and}
  \bibinfo{author}{\bibfnamefont{G.}~\bibnamefont{Mitchison}},
  \bibinfo{journal}{Nature} \textbf{\bibinfo{volume}{304}},
  \bibinfo{pages}{111} (\bibinfo{year}{1983}).

\bibitem[{\citenamefont{{Hopfield} et~al.}(1983)\citenamefont{{Hopfield},
  {Feinstein}, and {Palmer}}}]{Hopfield-1983}
\bibinfo{author}{\bibfnamefont{J.~J.} \bibnamefont{{Hopfield}}},
  \bibinfo{author}{\bibfnamefont{D.~I.} \bibnamefont{{Feinstein}}},
  \bibnamefont{and} \bibinfo{author}{\bibfnamefont{R.~G.}
  \bibnamefont{{Palmer}}}, \bibinfo{journal}{Nature}
  \textbf{\bibinfo{volume}{304}}, \bibinfo{pages}{158} (\bibinfo{year}{1983}).

\bibitem[{\citenamefont{Diekelmann and Born}(2010)}]{Sleep-2010}
\bibinfo{author}{\bibfnamefont{S.}~\bibnamefont{Diekelmann}} \bibnamefont{and}
  \bibinfo{author}{\bibfnamefont{J.}~\bibnamefont{Born}}, \bibinfo{journal}{Nat
  Rev Neurosci} \textbf{\bibinfo{volume}{11}}, \bibinfo{pages}{114}
  (\bibinfo{year}{2010}).

\bibitem[{\citenamefont{{Zhou} et~al.}(2020)\citenamefont{{Zhou}, {Lai}, {Bai},
  {Li}, {Zhao}, {Yang}, {Frank}, and {Gan}}}]{Zhou-2020rem}
\bibinfo{author}{\bibfnamefont{Y.}~\bibnamefont{{Zhou}}},
  \bibinfo{author}{\bibfnamefont{C.~S.~W.} \bibnamefont{{Lai}}},
  \bibinfo{author}{\bibfnamefont{Y.}~\bibnamefont{{Bai}}},
  \bibinfo{author}{\bibfnamefont{W.}~\bibnamefont{{Li}}},
  \bibinfo{author}{\bibfnamefont{R.}~\bibnamefont{{Zhao}}},
  \bibinfo{author}{\bibfnamefont{G.}~\bibnamefont{{Yang}}},
  \bibinfo{author}{\bibfnamefont{M.~G.} \bibnamefont{{Frank}}},
  \bibnamefont{and} \bibinfo{author}{\bibfnamefont{W.-B.} \bibnamefont{{Gan}}},
  \bibinfo{journal}{Nature Communications} \textbf{\bibinfo{volume}{11}},
  \bibinfo{pages}{4819} (\bibinfo{year}{2020}).

\bibitem[{\citenamefont{Hebb}(1949)}]{Hebb-1949}
\bibinfo{author}{\bibfnamefont{D.~O.} \bibnamefont{Hebb}},
  \emph{\bibinfo{title}{The organization of behavior}}
  (\bibinfo{publisher}{Wiley}, \bibinfo{address}{New York},
  \bibinfo{year}{1949}).

\bibitem[{\citenamefont{Cugliandolo and Tsodyks}(1994)}]{Tsodyks-1994}
\bibinfo{author}{\bibfnamefont{L.~F.} \bibnamefont{Cugliandolo}}
  \bibnamefont{and} \bibinfo{author}{\bibfnamefont{M.~V.}
  \bibnamefont{Tsodyks}}, \bibinfo{journal}{Journal of Physics A: Mathematical
  and General} \textbf{\bibinfo{volume}{27}}, \bibinfo{pages}{741}
  (\bibinfo{year}{1994}).

\bibitem[{\citenamefont{{Amit} et~al.}(1985)\citenamefont{{Amit}, {Gutfreund},
  and {Sompolinsky}}}]{Amit-1985}
\bibinfo{author}{\bibfnamefont{D.~J.} \bibnamefont{{Amit}}},
  \bibinfo{author}{\bibfnamefont{H.}~\bibnamefont{{Gutfreund}}},
  \bibnamefont{and}
  \bibinfo{author}{\bibfnamefont{H.}~\bibnamefont{{Sompolinsky}}},
  \bibinfo{journal}{Physical Review Letters} \textbf{\bibinfo{volume}{55}},
  \bibinfo{pages}{1530} (\bibinfo{year}{1985}).

\bibitem[{\citenamefont{{Kempter} et~al.}(1999)\citenamefont{{Kempter},
  {Gerstner}, and van {Hemmen}}}]{Hebbian-1999}
\bibinfo{author}{\bibfnamefont{R.}~\bibnamefont{{Kempter}}},
  \bibinfo{author}{\bibfnamefont{W.}~\bibnamefont{{Gerstner}}},
  \bibnamefont{and} \bibinfo{author}{\bibfnamefont{J.~L.} \bibnamefont{van
  {Hemmen}}}, \bibinfo{journal}{Physical Review E}
  \textbf{\bibinfo{volume}{59}}, \bibinfo{pages}{4498} (\bibinfo{year}{1999}).

\bibitem[{\citenamefont{Abbott and Nelson}(2000)}]{Abbott-2000}
\bibinfo{author}{\bibfnamefont{L.}~\bibnamefont{Abbott}} \bibnamefont{and}
  \bibinfo{author}{\bibfnamefont{S.}~\bibnamefont{Nelson}},
  \bibinfo{journal}{Nat Neurosci} \textbf{\bibinfo{volume}{3}},
  \bibinfo{pages}{1178} (\bibinfo{year}{2000}).

\bibitem[{\citenamefont{{Pfister} and {Gerstner}}(2006)}]{Tri-2006}
\bibinfo{author}{\bibfnamefont{J.-P.} \bibnamefont{{Pfister}}}
  \bibnamefont{and}
  \bibinfo{author}{\bibfnamefont{W.}~\bibnamefont{{Gerstner}}},
  \bibinfo{journal}{The Journal of Neuroscience} \textbf{\bibinfo{volume}{26}},
  \bibinfo{pages}{9673} (\bibinfo{year}{2006}).

\bibitem[{\citenamefont{{Gray}}(2006)}]{Gray-2005}
\bibinfo{author}{\bibfnamefont{R.~M.} \bibnamefont{{Gray}}},
  \bibinfo{journal}{Foundations and Trends in Communications and Information
  Theory} \textbf{\bibinfo{volume}{2}}, \bibinfo{pages}{155}
  (\bibinfo{year}{2006}).

\bibitem[{SM()}]{SM}
\bibinfo{note}{See supplemental material at http://... for technical details of
  replica method and mean-field dynamics, which includes
  Refs.~\cite{Gray-2005,Mezard-1987,Amit-1985}.}

\bibitem[{\citenamefont{{Hou} et~al.}(2019)\citenamefont{{Hou}, {Wong}, and
  {Huang}}}]{Huang-2019}
\bibinfo{author}{\bibfnamefont{T.}~\bibnamefont{{Hou}}},
  \bibinfo{author}{\bibfnamefont{K.~Y.~M.} \bibnamefont{{Wong}}},
  \bibnamefont{and} \bibinfo{author}{\bibfnamefont{H.}~\bibnamefont{{Huang}}},
  \bibinfo{journal}{Journal of Physics A: Mathematical and Theoretical}
  \textbf{\bibinfo{volume}{52}}, \bibinfo{pages}{414001}
  (\bibinfo{year}{2019}).

\bibitem[{\citenamefont{Hou and Huang}(2020)}]{Huang-2020}
\bibinfo{author}{\bibfnamefont{T.}~\bibnamefont{Hou}} \bibnamefont{and}
  \bibinfo{author}{\bibfnamefont{H.}~\bibnamefont{Huang}},
  \bibinfo{journal}{Phys. Rev. Lett.} \textbf{\bibinfo{volume}{124}},
  \bibinfo{pages}{248302} (\bibinfo{year}{2020}).

\bibitem[{\citenamefont{{Poe}}(2017)}]{Sleep-2017}
\bibinfo{author}{\bibfnamefont{G.~R.} \bibnamefont{{Poe}}},
  \bibinfo{journal}{The Journal of Neuroscience} \textbf{\bibinfo{volume}{37}},
  \bibinfo{pages}{464} (\bibinfo{year}{2017}).

\bibitem[{\citenamefont{Zhou et~al.}(2021)\citenamefont{Zhou, Jiang, Hou, Chen,
  Wong, and Huang}}]{Huang-2021b}
\bibinfo{author}{\bibfnamefont{J.}~\bibnamefont{Zhou}},
  \bibinfo{author}{\bibfnamefont{Z.}~\bibnamefont{Jiang}},
  \bibinfo{author}{\bibfnamefont{T.}~\bibnamefont{Hou}},
  \bibinfo{author}{\bibfnamefont{Z.}~\bibnamefont{Chen}},
  \bibinfo{author}{\bibfnamefont{K.~Y.~M.} \bibnamefont{Wong}},
  \bibnamefont{and} \bibinfo{author}{\bibfnamefont{H.}~\bibnamefont{Huang}},
  \bibinfo{journal}{to be submitted}  (\bibinfo{year}{2021}).

\bibitem[{\citenamefont{Fachechi et~al.}(2019)\citenamefont{Fachechi, Agliari,
  and Barra}}]{Barra-2019}
\bibinfo{author}{\bibfnamefont{A.}~\bibnamefont{Fachechi}},
  \bibinfo{author}{\bibfnamefont{E.}~\bibnamefont{Agliari}}, \bibnamefont{and}
  \bibinfo{author}{\bibfnamefont{A.}~\bibnamefont{Barra}},
  \bibinfo{journal}{Neural Networks} \textbf{\bibinfo{volume}{112}},
  \bibinfo{pages}{24} (\bibinfo{year}{2019}).

\bibitem[{\citenamefont{{Dotsenko} et~al.}(1991)\citenamefont{{Dotsenko},
  {Yarunin}, and {Dorotheyev}}}]{Dot-1991}
\bibinfo{author}{\bibfnamefont{V.~S.} \bibnamefont{{Dotsenko}}},
  \bibinfo{author}{\bibfnamefont{N.~D.} \bibnamefont{{Yarunin}}},
  \bibnamefont{and} \bibinfo{author}{\bibfnamefont{E.~A.}
  \bibnamefont{{Dorotheyev}}}, \bibinfo{journal}{Journal of Physics A}
  \textbf{\bibinfo{volume}{24}}, \bibinfo{pages}{2419} (\bibinfo{year}{1991}).

\bibitem[{\citenamefont{Nokura}(1998)}]{Nokura-1998}
\bibinfo{author}{\bibfnamefont{K.}~\bibnamefont{Nokura}}, \bibinfo{journal}{J.
  Phys. A: Math. Gen} \textbf{\bibinfo{volume}{31}}, \bibinfo{pages}{7447}
  (\bibinfo{year}{1998}).

\bibitem[{\citenamefont{{Fung} et~al.}(2010)\citenamefont{{Fung}, {Wong}, and
  {Wu}}}]{Fung-2010}
\bibinfo{author}{\bibfnamefont{C.~C.~A.} \bibnamefont{{Fung}}},
  \bibinfo{author}{\bibfnamefont{K.~Y.~M.} \bibnamefont{{Wong}}},
  \bibnamefont{and} \bibinfo{author}{\bibfnamefont{S.}~\bibnamefont{{Wu}}},
  \bibinfo{journal}{Neural Computation} \textbf{\bibinfo{volume}{22}},
  \bibinfo{pages}{752} (\bibinfo{year}{2010}).

\bibitem[{\citenamefont{Battista and Monasson}(2020)}]{Monasson-2020}
\bibinfo{author}{\bibfnamefont{A.}~\bibnamefont{Battista}} \bibnamefont{and}
  \bibinfo{author}{\bibfnamefont{R.}~\bibnamefont{Monasson}},
  \bibinfo{journal}{Phys. Rev. Lett.} \textbf{\bibinfo{volume}{124}},
  \bibinfo{pages}{048302} (\bibinfo{year}{2020}).

\bibitem[{\citenamefont{Mezard et~al.}(1987)\citenamefont{Mezard, Parisi, and
  Virasoro}}]{Mezard-1987}
\bibinfo{author}{\bibfnamefont{M.}~\bibnamefont{Mezard}},
  \bibinfo{author}{\bibfnamefont{G.}~\bibnamefont{Parisi}}, \bibnamefont{and}
  \bibinfo{author}{\bibfnamefont{M.~A.} \bibnamefont{Virasoro}},
  \emph{\bibinfo{title}{Spin Glass Theory and Beyond}}
  (\bibinfo{publisher}{World Scientific}, \bibinfo{address}{Singapore},
  \bibinfo{year}{1987}).

\end{thebibliography}

%%%%%%%%%%%%%%%%%%%%%%%%%%%%%%%%%%%%%%%%%%%%%%%%%%%%%%%%%%%%%%%%%%%%%

\end{document}